\documentclass[amssymb,11pt]{article}
\usepackage[pdftex,colorlinks=true,urlcolor=black,filecolor=black,linkcolor=black,
            pdftitle={Non-metric Quantum Cosmology},
            pdfauthor={Mark D. Roberts},
            pdfsubject={cosmology},
            pdfkeywords={cosmology},
            pagebackref,pdfpagemode=None,bookmarksopen=true]{hyperref}
\usepackage{amssymb}
\newcommand{\bc}{\begin{center}}
\newcommand{\ec}{\end{center}}
\newcommand{\be}{\begin{equation}}
\newcommand{\ee}{\end{equation}}
\newcommand{\ber}{\begin{eqnarray}}
\newcommand{\ear}{\end{eqnarray}}

\newcommand{\ph}{\phi}

\begin{document}
\title{Non-metric Quantum Cosmology}
\author{Mark D. Roberts}
\date{$24^{th}$ of November 2016}
\maketitle
\abstract{Scalar-tensor theory has arbitrary functions of the scalar field in front of the
geometric and scalar terms in the Lagrangain.   The extent to which these arbitrary functions
appear in the Wheeler-deWitt wavefunction of mini-super Robertson-Walker spacetimes is discussed.
The function of the scalar field in front of the Einstein-Hilbert action allows a current
to be constructed which suggests transfer from matter to geometry of aspects of the wavefunction.}
\tableofcontents
\section{Introduction}\label{intro}
Scalar-tensor theory is a common variant of general relativity with arbitrary functions
of the scalar field present,  here the extent to which this arbitrariness follows through
to quantum theory is investigated.  Scalar-tensor quantum cosmology has been previously
studied
by Pimentel and Mora \cite{pm},
by Fabris,  Pinto-Neto and Velasco \cite{fpv}
by Xu,  Harko,  and Liang \cite{xhl}.
The line element is taken to be of general Robertson-Walker form
\begin{equation}
\label{rw}
ds^2=-N(t)^2dt^2+a(t)^2\left\{\frac{dr^2}{1-kr^2}+r^2(d\theta^2+\sin(\theta)^2d\phi^2)\right\},
\end{equation}
and the Hamiltonians considered here are taken to be of the form
\begin{equation}
\label{hamiltonian}
H=\frac{N}{2}G^{AB}\Pi_A\Pi_b+NU(q),
\end{equation}
where $A,B\dots$ range over the scale factor and the scalar field.
Quantization is achieved via
\begin{equation}
\label{quant}
\Pi_A\rightarrow -i\hbar\nabla_A,
\end{equation}
applying (\ref{quant}) to (\ref{hamiltonian}) gived the Wheeler-deWitt equation
\begin{equation}
\label{wdw}
H\psi=-\hbar^2\Box\psi+NU\psi.
\end{equation}
Conventions used include signature $-+++$,
potential $V(\phi)$ for the scalar field and potential $U(q)$ for the Wheeler-deWitt potential.
\section{Cosmological constant}\label{cc}
The treatment of the next two sections is closest to that in Keifer \cite{keifer}.
The field equations equations are taken to be
\begin{equation}
\label{fcc}
GC_{ab}=G_{ab}+\Lambda g_{ab},
\end{equation}
subect to the geometry the Friedman equation is
\begin{equation}
\label{frie}
f=-GC^t_{.t}=\frac{1}{N^2a^2}\left(3kN^2+3\dot{a}^2-\Lambda a^2N^2\right),
\end{equation}
and the spatial equation is
\begin{equation}
\label{spatial}
S=+GC^r_{.r}=\frac{1}{a^2N^3}\left(-kN^3+\dot{a}^2N-2a\dot{a}\dot{N}-2a\ddot{a}N+\Lambda a^2N^3\right).
\end{equation}
The velocity Hamiltonian equation is
\begin{equation}
\label{velh}
\dot{q}_A=\frac{\delta H}{\delta \Pi_A}=NG^{AB}\Pi_B,
\end{equation}
from which $G$ has one component $q=a$ and
\begin{equation}
\label{cmom}
G^{AB}=-\frac{1}{a},~~~
\Pi_=-\frac{a\dot{a}}{N},
\end{equation}
substituting into (\ref{hamiltonian}) gives
\begin{equation}
\label{hf}
H=-\frac{Na^3F}{6},
\end{equation}
provided
\begin{equation}
\label{cu}
U=\frac{1}{2}\left(-ka+\frac{\Lambda}{3}a^3\right).
\end{equation}
For the momentum Hamiltonian equation
\begin{equation}
\label{ccmom}
\frac{1}{2}a^2NS=\frac{\delta H}{\delta q}+\dot{\Pi}.
\end{equation}

Thus the Hamiltonian constraint is related to the Friedman equation (\ref{hf}),
the $\dot{q}$ Hamiltonian equation is the relationship between $q$ and $\Pi$ (\ref{velh}),
the $\dot{\Pi}$ Hamiltonian equation is the spatial equation (\ref{ccmom}).

Quantization of the Hamiltonian constraint gives the Wheeler-deWitt equation (\ref{wdw})
\begin{equation}
\label{wdwcc}
\hbar^2\psi_{aa}+a^2\left(-k+\frac{\Lambda}{3}a^2\right)=0,
\end{equation}
for $\Lambda=0$ (\ref{wdwcc}) has solution
\begin{equation}
\label{psik}
\psi=C_1\sqrt{a}BessJ\left(\frac{1}{4},\frac{\sqrt{-k}a^2}{2\hbar}\right)
    +C_2\sqrt{a}BessY\left(\frac{1}{4},\frac{\sqrt{-k}a^2}{2\hbar}\right),
\end{equation}
for $k-0$ (\ref{wdwcc}) has solution
\begin{equation}
\label{psil}
\psi=C_1\sqrt{a}BessJ\left(\frac{1}{6},\frac{\sqrt{\Lambda}a^2}{3\hbar}\right)
    +C_2\sqrt{a}BessY\left(\frac{1}{6},\frac{\sqrt{\Lambda}a^2}{3\hbar}\right),
\end{equation}
where $BessJ,~BessY$ are Bessel functions.
No more solutions have been found for $\psi$.
If one defines a principle function
\begin{equation}
\label{principle}
\psi=C\exp\left(\frac{iS}{\hbar}\right),
\end{equation}
then there seem to be no relationship between $S_a$ and the momentum in particular because
there is no velocity $\dot{a}$ in the resulting equations.
\section{Massive scalar field}\label{msf}
To extend the treatment of the previous section \S\ref{cc}
to include a massive scalar field with stress
\begin{equation}
\label{msfstress}
T_{ab}=6\phi_a\phi_b-3g_{ab}\left(\phi_c\phi_dg^{cd}+m^2\ph^2\right),
\end{equation}
first
\begin{equation}
\label{newgc}
GC_{ab}=GC_{old~ab}-T_{ab},
\end{equation}
with an additional momenta
\begin{equation}
\label{msfmom}
\Pi_\phi=\frac{a^3\dot{\phi}}{N},
\end{equation}
the mini-metric is
\begin{equation}
\label{msfmm}
G_{AB}=\left(
\begin{array}{cc}
-a & 0\\
0  &a^3
\end{array}\right),
\end{equation}
and the potential is
\begin{equation}
\label{msfpot}
U=U_{old}+\frac{1}{2}m^2a^3\phi^3,
\end{equation}
now the field equations and Euler equation are
\begin{equation}
\label{msffld}
F=F_{old}-\frac{3\dot{\phi}}{n^2}-3m^2\phi^2,~
S=S_{old}+\frac{3\dot{\phi}}{n^2}-3m^2\phi^2,~
E=-\frac{1}{N^2}\left(\ddot{\phi}+\frac{3\dot{\phi}\dot{a}}{a}-\dot{\phi}\dot{N}\right)-m\phi.
\end{equation}
The Hamiltonian equations are as before with the addition
\begin{equation}
\label{phimom}
-a^3NE=\frac{\delta H}{\delta \phi}+\dot{\Pi_\phi}.
\end{equation}
Quantization gives the Wheeler-deWitt equation (\ref{wdw})
\begin{equation}
\label{sfwdw}
H\psi=\frac{N\hbar^2}{2a^3}\left(a^2\psi_{aa}+a\psi_a-\psi_{\phi\phi}\right)+NU\psi,
\end{equation}
transforming to linear null fields
\begin{equation}
\label{nullsf}
2u\equiv \ln(a)+\phi,~~~
2v\equiv \ln(a)-\phi,
\end{equation}
the Wheeler-deWitt equation (\ref{sfwdw}) becomes
\begin{equation}
\label{whwsfnull}
\frac{\hbar^2}{\phi}\phi_{uv}-k\exp(4(u+v))+\frac{\Lambda}{3}\exp(6(u+v))+m^2(u-v)\exp(6(u+v)).
\end{equation}
A solution for $\lambda=0$ is linear combinations of $C_+$ and $C_-$ parts is
\begin{equation}
\label{sfcpm}
\psi=C_\pm\exp\left(\pm\frac{\sqrt{k}}{4\hbar}(e^{4u}+e^{4v})\right).
\end{equation}
A solution for $m=0$ is linear combinations os $C_+$ and $C_-$ parts is
\begin{equation}
\label{sfcpl}
\psi=C_\pm\exp\left(\pm\frac{\sqrt{-\Lambda/3}}{6\hbar}(e^{6u}+e^{6v})\right).
\end{equation}.
\section{Scalar-tensor theory}\label{stt}
The scalar-tensor action is taken to be
\begin{equation}
\label{action}
S=\int dx^4 \sqrt{-g}\left\{A(\phi)R-B(\phi)\phi_c^2-V(\phi)\right\},
\end{equation}
Palatini varying the action and then replacing the non-metric connection with a Christoffel
connection gives \cite{mdrwy}
\begin{equation}
\label{nmstress}
8\pi\kappa^2T_{ab}=AG_{ab}-A'\phi_{ab}-\left(B+A"-\frac{3A'^2}{2A}\right)\phi_a\phi_b
+\frac{1}{2}\left[2A'\Box\phi+\left(B+2A"-\frac{3A'^2}{2A}\right)\phi_c^2+V\right],
\end{equation}
subject to $N=1$ Robertson-Walker geometry (\ref{rw})
\begin{equation}
\label{hp}
2H=-\frac{a^3F}{3}=\frac{8\pi\kappa^2}{3}a^3T^t_{.t}=
\frac{a}{12A}\left[-12A^2\dot{a}^2-12A'aA\dot{\phi}\dot{a}+a^2(2AB-3a'^2)\dot{\phi}^2\right]
+U(q)
\end{equation}
making the cosmological constant explicit the potential is
\begin{equation}
\label{pot}
U(q)=-kAa+\frac{\Lambda}{3}a^3+\frac{a^3}{6}V(\phi),
\end{equation}
(\ref{hp}) can be diagonalized by defining
\begin{equation}
\label{defal}
\alpha\equiv\sqrt{A}a,
\end{equation}
then
\begin{equation}
\label{dh}
2H=-\frac{\alpha\dot{\alpha}^2}{\sqrt{A}}+\frac{B\alpha^3\dot{\phi}^2}{6A^\frac{3}{2}}+U(q),
\end{equation}
the momenta are
\begin{equation}
\label{nmmm}
\Pi_\alpha=-\frac{\alpha\dot{\alpha}}{\sqrt{A}},~~~
\Pi_\phi=\frac{B\alpha^3\dot{\phi}}{6A^\frac{3}{2}},
\end{equation}
using (\ref{nmmm}) the Hamiltonian (\ref{dh}) can be written as
\begin{equation}
\label{nmpih}
2H=-\frac{\sqrt{A}}{\alpha}\Pi_\alpha^2+\frac{6A^\frac{3}{2}}{B\alpha^3}\Pi_\phi^2+U(q),
\end{equation}
from which can be read off the mini-metric
\begin{equation}
\label{mmgeom}
G^{AB}=-\frac{\sqrt{A}}{\alpha}
\left(
\begin{array}{cc}
-1 & 0\\
0 & \frac{6A}{B\alpha^2}
\end{array}
\right),~~~
G_{AB}=-\frac{\alpha}{\sqrt{A}}
\left(
\begin{array}{cc}
-1 & 0\\
0 & \frac{B\alpha^2}{6A}
\end{array}
\right),
\end{equation}
and mini-determinant and mini-curvature
\begin{equation}
\label{nmgdetcur}
\sqrt{-G}=\frac{\alpha^2}{A}\sqrt{\frac{B}{6}},~~~~~~~
R_G=\frac{3}{\alpha^3A^2B^\frac{5}{2}}\left(\frac{A_\phi}{\sqrt{AB}}\right)_\phi.
\end{equation}

Quantization is implemented as before,
assuming no $\psi R_G$ term and
using the variable $\beta=\ln(\alpha)$
gives Wheeler-deWitt equation
\begin{equation}
\label{nmwdw}
\frac{2e^{3\beta}H\psi}{\hbar^2\sqrt{A}}=\psi_{\beta\beta}
-6\sqrt{\frac{A}{B}}\left(\sqrt{\frac{A}{B}}\psi_\phi\right)_\phi
+\frac{Ue^{3\beta}\psi}{\hbar^2\sqrt{A}}=0.
\end{equation}
First consider the $U=0$ separation of variables case,
then using $C(\phi)=\sqrt{A(\phi)/B(\phi)}$
one ends up with the ordinary differential equation $6C(\phi)[C(\phi)\Phi_\phi]_\phi-\Phi=0$,
so one can choose a suitably differentiable $\Phi$ then there is a $C(\phi)$
to satisfy it and in this sense the arbitrariness of scalar-tensor theory follows through.
Now consider the case where the potential is just the cosmological constant then
\begin{equation}
\label{nmbccwdw}
\psi_{\beta\beta}-6\sqrt{\frac{A}{B}}\left(\sqrt{\frac{A}{B}}\psi_\phi\right)_\phi
+\frac{\Lambda e^{6\beta}\psi}{3\hbar^2A^2}=0,
\end{equation}
now fix the arbitrary functions with
\begin{equation}
\label{defk}
A=C_Ae^{k\phi},~~~
B=C_Be^{k\phi},~~~
\frac{A}{B}=\frac{C_A}{C_B}=\frac{1}{6},~~~
k\ne\pm3,
\end{equation}
this choice has vanishing mini-curvature $R_G=0$,
$k$ would normally be close to $1$ so as to be close to existing theory so the
condition $k\ne3$ is not physically restrictive,
now (\ref{nmbccwdw}) becomes
\begin{equation}
\label{whwpp}
\psi_{\beta\beta}-\psi_{\phi\phi}+4\ell\exp(2(3\beta-k\psi))\psi=0,~~~
\ell\equiv\frac{\Lambda}{12\hbar^2C_A^2},
\end{equation}
defining null coordinates $2u=\beta-\phi,~2v=\beta+\phi$ (\ref{whwpp}) becomes
\begin{equation}
\psi_{uv}+\ell\exp\left(2(3+k)u+2(3-k)v\right)\psi=0,~~~
\end{equation}
which has solution
\begin{equation}
\label{nmsol}
\psi=C\exp \frac{i}{4\hbar C_A}\sqrt{\frac{\Lambda}{3(3-k)(3+k)}}\left[e^{2(3+k)u}+e^{2(3-k)v}\right].
\end{equation}
The current is defined as
\begin{equation}
\label{currentdef}
j_a=i(\psi\psi^*_a-\psi^*\psi_a)
\end{equation}
for (\ref{nmsol}) this is
\begin{equation}
\label{nmcur}
j_a=2C^2\left[\sqrt{\frac{(3+k)\ell}{3-k}},\sqrt{\frac{(3-k)\ell}{3+k}}\right],
\end{equation}
which has size
\begin{equation}
\label{size}
j_aj^a=-\frac{\Lambda C^2 A^{-\frac{3}{2}}\alpha^3}{3\hbar^2},
\end{equation}
as (\ref{size}) is negative the current is timelike perhaps indicating transfer
of aspects of the wavefunction from matter to geometry.
\section{Conclusion}\label{conc}
Given the complexity of the stress (\ref{nmstress}),
the corresponding Hamiltonian (\ref{nmpih}) turns out to be surprisingly simple
and in general has non-vanishing mini-Ricci scalar (\ref{nmgdetcur}).
For vanishing Wheeler-deWitt potential $U=0$ and assuming separation of variables,
to a certain extent the arbitrariness of
scalar-tensor theory follows through to quantum cosmology.
That the wavefunction current (\ref{nmcur}) is timelike could be considered an indication
that an aspect of the wavefunction is moving from matter to geometry however there are
at least three problems with this.
Firstly there seems to be no indication of transfer in the Born density $\rho=\psi\psi^*$
which is just a constant.
Secondly there could be other solutions to (\ref{nmwdw}) with different properties.
Thirdly there is no indication so far as to what happens in the corresponding classical
picture in particular whether energy is transferred.
This mechanism can be thought of as a possible origin of spacetime: for others see \cite{mdr42}


\begin{thebibliography}{99}


\bibitem{fpv}
Julio C. Fabris,  Nelson Pinto-Neto,  A.F. Velasco,
Quantum Cosmology in Scalar-tensor theories with non-minimal coupling.\\
\href{http://arXiv.org/abs/gr-qc/990311}{\tt \color{blue}990311v1}.

\bibitem{keifer}
Claus Kiefer,
Quantum Gravity,
Oxford University Press 2004.

\bibitem{pm}
L.O. Pimentel, C. Mora,
Quantum Cosmology in some scalar-tensor theories.\\
\href{http://arXiv.org/abs/gr-qc/990311}{\tt \color{blue}980302561}.

\bibitem{mdr42}
Mark D. Roberts,
The Relative Motion of Membranes.\\
{\it Central European Journal of Physics}{\bf 8}(2010)\href{http://link.springer.com/article/10.2478}{915-919},
\href{http://www.slac.stanford.edu/spires/find/hep/www?eprint=gr-qc/0404094}{\color{red}spires},
\href{http://dx.doi.org/10.2478/s11534-010-0022-z}{\color{yellow}doi},
\href{http://arXiv.org/abs/gr-qc/0404094}{\tt \color{blue}gr-qc/0404094}.

\bibitem{mdrwy}
Mark D. Roberts,
Is spacetime non-metric?
\href{http://arXiv.org/abs/0706.4043}{\tt \color{blue}0706.4043},
\href{http://www.slac.stanford.edu/spires/find/hep/www?eprint=0706.4043}{\color{red}spires}.

\bibitem{xhl}
Min-Xing Xu,  Tiberiu Harko,  Shi-Dong Liang,
Quantum Cosmology of f(R,T) gravity.\\
\href{http://arXiv.org/abs/1608.00113}{\tt \color{blue}1608.00113}.

\end{thebibliography}
\end{document}